\documentclass[%
 reprint,
 amsmath,amssymb,
 aps,
]{revtex4-2}

\usepackage{graphicx}
\usepackage{dcolumn}
\usepackage{bm}

\begin{document}

\preprint{APS/123-QED}

\title{\textit{Ab initio} modelling of spin relaxation lengths in disordered graphene nanoribbons}

\author{Wudmir Y. Rojas}
 \affiliation{Grupo de Investigaci\'on en F\'isica, Universidad San Ignacio de Loyola,  Av. la Fontana 550, La Molina 00012, Lima, Peru.}
\affiliation{Centro de Ci\^encias Naturais e Humanas, Universidade Federal do ABC, Santo Andr\'e, Brazil.}
\author{Cesar E. P. Villegas}%
\affiliation{Departamento de Ciencias, Universidad Privada del Norte, Av. Alfredo Mendiola 6062, Los Olivos, Lima, Peru.}%
\author{Alexandre R. Rocha}%
\affiliation{Instituto de F\'{\i}sica Te\'{o}rica, Universidade Estadual Paulista (UNESP), Rua Dr. Bento T. Ferraz, 271, S\~{a}o Paulo, SP 01140-070, Brazil.}%




\date{\today}

\begin{abstract}
The spin-dependent transport properties of armchair graphene nanoribbons in the presence of extrinsic spin-orbit coupling induced by a random distribution of Nickel adatoms is studied. By combining a recursive Green's function formalism with density functional theory, we explore the influence of ribbon length and metal adatom concentration on the conductance. At a given length, we observed a significant  enhancement of the spin-flip channel around resonances and at energies right above the Fermi level. We also estimate the spin-relaxation length, finding values on the order of tens of micrometers at low Ni adatom concentrations. 
This study is conducted at singular ribbon lengths entirely from fully \emph{ab-initio} methods, providing  indirectly evidence that the Dyakonov-Perel spin relaxation mechanism might be the dominant at low concentrations as well as the observation of oscillations in the spin-polarization. 
\end{abstract}

\maketitle


\section{Introduction}

The design of spintronic devices, \emph{i.e.}, the use of the spin degree of freedom of electrons for applications, is a field full of promise, from both fundamental and technological points of view. \cite{dassarma2004} Since spintronic devices rely on the  controlled production of a spin current, and
its subsequent output signal detection, the correct understanding of spin-scattering mechanisms within
a device is crucial. In this regard, this phenomena has been explored both theoretically and experimentally in metals, semiconductors, and molecules.\cite{fert2008,wu2010}

Most recently, graphene-based spintronics has attracted considerable attention by virtue of its micrometer scale spin coherence length measured at room temperature.\cite{avsar2011,han2011} While, the weak intrinsic spin-orbit coupling (SOC) and low hyperfine structure in graphene is preferred for spin transport, it has been experimentally shown that, regardless of the growth substrate\cite{han2011} or even when suspended graphene is obtained,\cite{guimaraes2012} the spin-relaxation lengths do not appear to significantly change. In this regard, a number of extrinsic mechanisms including, corrugation, \cite{jeong2011} impurities,\cite{pi2010} disorder and adatoms have been addressed to elucidate the puzzling discrepancies of spin diffusion in graphene.\cite{gmitra2013,kochan2014,cresti2014,vantuan2016}


Graphene nanoribbons (GNR) are highly attractive for new class of electronics, optoelectronics and spin transport devices, indeed, GNRs are well-suited for spintronic applications.\cite{gnr-opto,villegas, gnr-trans1} At the same time, by considering that GNRs present more chemically reactive edges,\cite{jiang2007,bellunato2016} one might expect that metal adsorption and functionalization can, in principle, lead to considerable larger binding energies when compared with pristine graphene,\cite{Ataca2011,MANADE2015,Brito2010} which certainly would promotes extrinsic sources of SOC. 

This turns out to be more crucial, when one has to consider current bottom-up synthesis methods that leads to the realization of well defined edges with semiconducting GNRs,\cite{gnr1} and their subsequent device applications. Regardless of the extensive works addressing the effects of spin polarization,\cite{kim2008,dutta2010} impurities and adatoms,\cite{Rzeszotarski2017,ganguly2017,cocchi2010,rigo2009,biel2009,Krychowski2014} there yet is scarce literature concerning the spin-relaxation processes in GNRs,\cite{Bundesmann2015,Chaghazardi2016,droth2013,rao2012,wilhelm2015,salimath2014} and there is still a lack of comprehension of spin-lifetimes in GNRs.\cite{Son2006} 

\begin{figure*}[ht]
\centering
\includegraphics[scale=0.55]{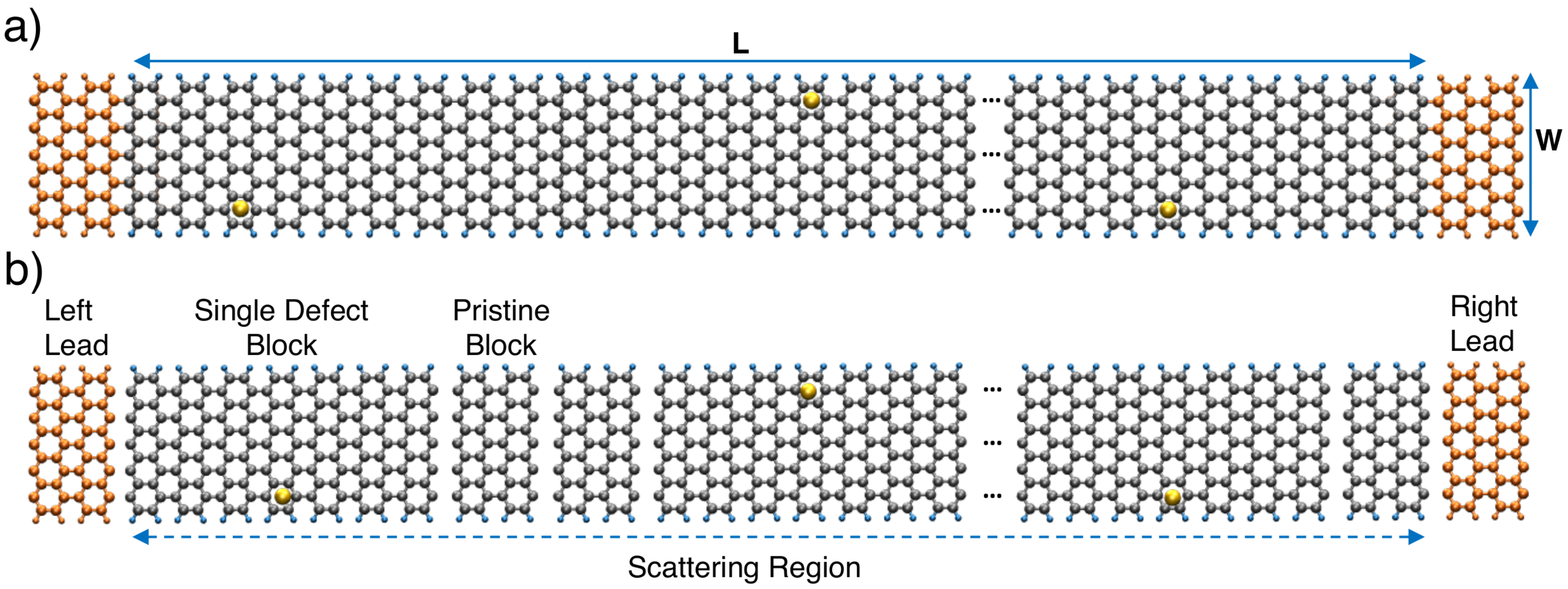}
\caption{Schematic representation of an 11-armchair graphene nanoribbon of length $L$ containing a large number of randomly distributed impurities. (a) Ni adatoms in a disordered armchair graphene nanoribbon. (b) Geometrical features of a disordered armchair graphene nanoribbon.}\label{fig1}
\end{figure*}

As a matter of fact, spin dynamics has been studied experimentally in GNRs using electron spin  resonance spectroscopy, finding relaxation times that are strongly temperature-dependent with values on the order of microseconds.\cite{rao2012} Theoretically, Chaghazardi \emph{et al.}, using
a tight-binding model studied an armchair GNR (tens of nanometers in length) in the presence of surface roughness, estimating spin-diffusion lengths ranging from $2 \ \mu m$ to $300\ \mu m$,  which corresponds to spin-lifetimes on the order of nanoseconds.\cite{Chaghazardi2016} Moreover, Salimath and Ghosh, performed Monte-Carlo simulations of long armchair GNRs ($\approx 5\ \mu m$ length), finding spin-relaxation lengths on the order of $1-4$ micrometers.\cite{salimath2014} As these calculations were performed on model Hamiltonians, this collection of results clearly suggest the need of first-principles studies to elucidate key mechanisms regarding spin-diffusion lengths in armchair GNRs (AGNR).

In this work we use density functional theory (DFT) combined with recursive Green's functions methods---taking into account spin-orbit coupling---to carry out spin-dependent transport in Ni-doped disordered AGNR with lengths of up to $\approx$ $2 \ \mu m$.  Our findings show spin-relaxation
lengths on the order of tens micrometers. Moreover, the existence of resonance centers in the electron transmission enhances the spin-flipping scattering paths, which indicate that those can act as \emph{spin hot spots}.\cite{kochan2014} Our results also imply that at lower adatom concentrations, the dominant spin relaxation mechanism resembles the one proposed by Dyakonov-Perel (DP).\cite{DP1972}

\section{Methodology}
The simulated structure is an 11-AGNR along the $z$-axis in which Ni atoms are adsorbed on top of carbon's rings at the edges as shown in Figure \ref{fig1}(a-b). As previously reported, this configuration is the most stable site for adsorption.\cite{cocchi2010,rojas2018} The system's electronic properties were computed within the density functional theory (DFT),\cite{HK,KS} as implemented in SIESTA,\cite{siesta} with an energy cut-off of 350 Ry and using a $k$-space grid of 1 $\times$ 1 $\times$ 20 in the Monkhorst-Pack scheme along the $z$-direction. The local spin-density approximation (LDA) for the exchange correlation functional, in addition to the double-$\xi$ basis set with polarization orbitals is employed. The atomic coordinates were optimized until reaching a force threshold of 0.02 eV/\AA. 

For the spin-dependent transport properties, we adopted the procedure proposed by Caroli \emph{et al}., \cite{caroli} in which the system is divided into three regions, namely two semi-infinite electrodes ($L/R$) and a central scattering region ($S$), as depicted in Figure \ref{fig1}a. The retarded Green's function at a given energy $E$ is thus given by

\begin{equation}\label{1}
{G}^{R}_{S}(E,L)= \left[E \times S_{S} - H^{}_{S}(L) - \Sigma^{}_{L}(E) - \Sigma^{}_{R}(E) \right]^{-1},
\end{equation}
where $H^{}_{S}$ and $S_{S}$ are the Hamiltonian and overlap matrices for the scattering region, and $\Sigma^{}_L(E)$ ($\Sigma^{}_R(E))$ represent the self-energy for the left- (right-) hand-side electrode which account for the coupling to the leads. From this, the Landauer-B\"uttiker formula \cite{buti-lan} is used to calculate the transmission probability of the system 
\begin{equation}\label{2}
T(E) =  Tr[\Gamma^{}_{L} G^{R^{\dagger}}_{S} \Gamma_{R}^{} G^{R}_{S}] ~,
\end{equation}
where $ \Gamma^{}_{L/R}(E) = i[\Sigma^{}_{L/R}(E) - \Sigma^{}_{L/R}(E)^{\dagger}]  $ are the coupling matrices that represent the rates at which electrons are  scattered into (or out of) the ribbon. Without SOC, the spin space is block diagonal and all quantities can be calculated for each spin independently.

In the presence of SOC, but considering that SOC is present only in the scattering region, the full Spin-Hamiltonian must be considered, and the transmission coefficient comprises the contribution of spin degree of freedom in the form
\begin{equation}\label{1.74}
T(E)=  T^{\uparrow\uparrow}(E,L) + T^{\uparrow\downarrow}(E,L) + T^{\downarrow\uparrow}(E,L) + T^{\downarrow\downarrow}(E,L) ~,
\end{equation}
where the relevant transmission probabilities are given by  
\begin{eqnarray}\label{1.73}
T_{sc} = T^{\sigma\sigma}(E,L)=Tr[\Gamma_{L}^{\sigma\sigma} (G^{\sigma\sigma}_{S}(E,L))^{\dagger} \Gamma_{R}^{\sigma\sigma} G^{\sigma\sigma}_{S}(E,L)] ~,\\
T_{sf} = T^{\sigma\sigma'}(E,L)=Tr[\Gamma_{L}^{\sigma\sigma} (G^{\sigma\sigma'}_{S}(E,L))^{\dagger}\Gamma_{R}^{\sigma'\sigma'} G^{\sigma\sigma'}_{S}(E,L)] ~
\end{eqnarray}

here $T^{\sigma\sigma'}(E,L)$ represents the probability that carriers with up-spin (down-spin) will be out-scattered with down-spin (up-spin), \emph{i.e.}, a signature that spin-flip processes occur in the channel due to SOC. Accordingly, $T^{\sigma\sigma}(E,L)$ corresponds to the spin-conserved terms, \emph{i.e.}, incoming electrons are out-scattered with the same spin.

In principle, the procedure described above can be used in disordered systems, but dealing with a system with hundreds of nanometers in length greatly increased the scattered region, thus hindering the numerical inversion of equation (\ref{1}). To overcome this issue, we use the recursive Green's function method for describing systems with a large number of randomly distributed impurities.\cite{amorim2013,rocha2007,rocha2010} To this aim, one realization of a Ni-doped disordered GNR is cleaved up to smaller segments that can be either a fragment of pristine AGNR or a ribbon with an adatom (see Figure \ref{fig1}b). Thus, one is able to obtain the Hamiltonian and overlap matrices of each building block using ground state DFT. Subsequently, the Hamiltonian, $H^{\sigma}_{S}$, is reconstructed by randomly arranging the blocks for a given length and defect concentration. Note that the coupling between blocks is assumed to be equal to that corresponding to a pristine AGNR by matching the potential at the edges of each segment. This procedure enables us to simulate unprecedented disordered AGNR by joining segments with randomly adsorbed atoms with the pristine ones, which span from approximately 0.1 to 2 $\mu m$ as depicted in Figure \ref{fig1} (a-b). 

Since we are interested in studying spin-flipping events, we compute the spin-polarization of transmitted electrons,
\begin{equation}\label{2}
P(E,L) = \frac{T^{\sigma\sigma}(E,L)-T^{\sigma\sigma'}(E,L)}{T^{\sigma\sigma}(E,L)+T^{\sigma\sigma'}(E,L)} ~,
\end{equation}
which is a suitable observable to assess in the study of spin-diffusion processes.

As we also analyzed the transmission dependence with adatom's concentration, it is worth to mention that the concentration of adatoms in the AGNR can be tuned by modifying the relative number of building blocks containing Ni atoms (see Supplementary information for details). Finally, in order to obtain statistically representative values for transmission, we have calculated up to $100$ different realizations and presented an average for each concentration and length of the AGNRs.

\section{Results}
\subsection{Spin channel finite size effects}
\begin{figure}[t]
\centering
\includegraphics[scale=0.38]{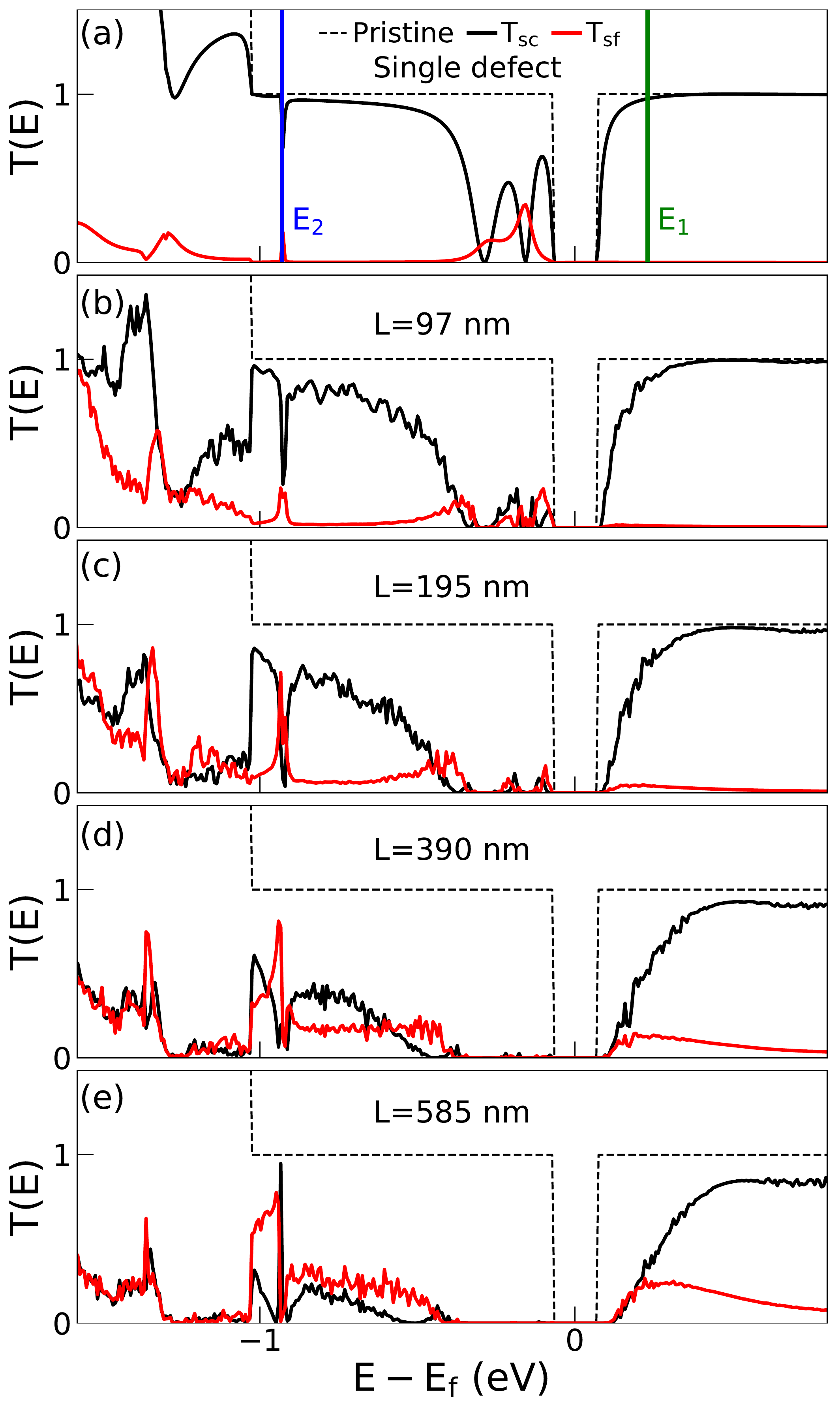}
\caption{Transmission probability for spin-conserved (black lines) and spin-flip (red lines) for different ribbon lengths ($L$) at a Ni concentration
of $\approx 0.5\%$. The single defect case is included for the sake of clarity of the resonance observed in the valence band. Dashed lines indicate the transmission for a pristine AGNR, while the vertical lines highlight energies $E_{1}$ and $E_{2}$ taken for the relaxation length analysis.}\label{fig2}
\end{figure}

To analyze the effect of ribbon length on the transmission, the adatom concentration is fixed at $\approx 0.5 \%$. Figure \ref{fig2}(b-e) depicts the spin transmission for four different lengths up to $\approx 600\ nm$. For all cases, at the conduction band, the spin-conserved transmission presents
similar features that slightly decrease with ribbon length. Conversely, the spin-flip transmission significantly increases with the ribbon length. Note that this effect follows an opposite trend to the one observed by Chaghazardi \emph{et al.} in AGNRs.\cite{Chaghazardi2016} We argue that such discrepancy might be related to the fact that the authors studied nanoribbons whose lengths are smaller than $100\ nm$. 

Regarding the trends of transmission at the valence band, we observed that the spin-conserved transmission significantly decreases for energies around the Fermi level while the spin-flip channel is rapidly enhanced for energies below $-0.4\ eV$. In contrast, it is suppressed for values
above $-0.4\ eV$.  In particular, at energies around the resonance ($E\approx -0.93\ eV$) one can clearly observe, in all cases, the increase of the spin-flip transmission probability. This effect is consistent with previous reports suggesting that the presence of resonance centers in the
electron transmission might act as \emph{spin hot spots}, \emph{i.e.},\cite{kochan2014} spin-flipping scattering rates are enhanced provided that exchange interaction time is comparable with the spin-precession time as we will discuss later.

\subsection{Spin relaxation length}
\begin{figure}[ht]
\centering
\includegraphics[scale=0.36]{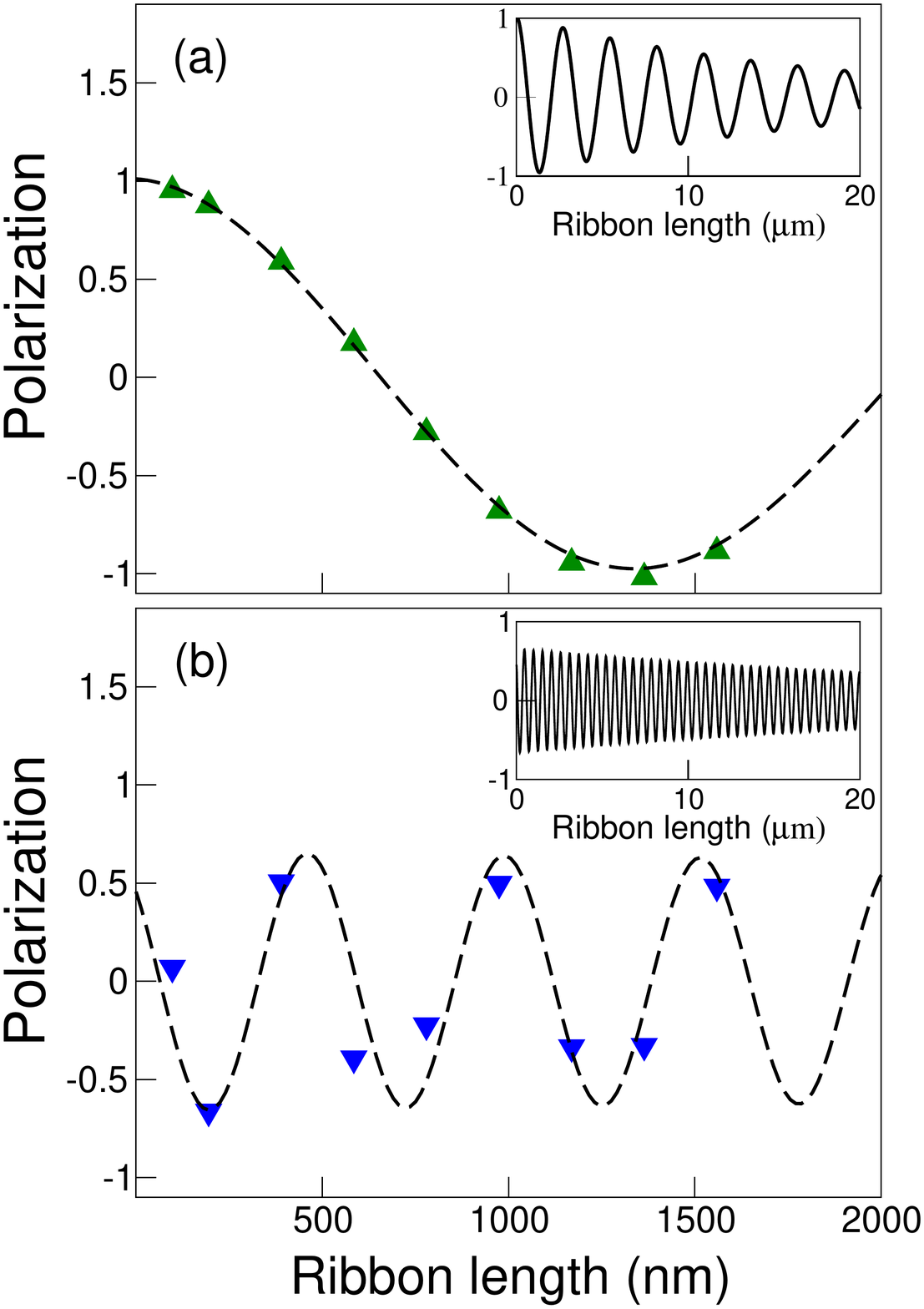}
\caption{Spin polarization as a function of the ribbon length for a given energy: (a) $E_{1}$, at the conduction band and (b) $E_{2}$, at the valence band, $i.e.$, energies around the resonance. The inset shows a zoom out of the polarization's oscillations.}\label{fig3}
\end{figure}

Turning to the spin-diffusion length in the AGNRs. In Figure \ref{fig3}(a-b) we present the spin polarization as a function of the ribbon length for two energies (i) $E_{1}= 0.23 \ eV$ and (ii) $E_2 \approx -0.93 \ eV$, corresponding to the CB and VB, respectively. These points were chosen to depict the mechanisms associated to the conduction band right above the Fermi energy and around the resonance. Indeed, $E_2$ represented an averaged value of energies within the resonance.  The observed oscillations are fingerprints of the spin precession that is induced due to the Rashba SOC.\cite{tuan2016} In fact, such oscillatory behavior has also been observed in a two-dimensional electron gas and related directly to quantum mechanical effects.\cite{pareek2002} In addition, by fitting the DFT polarization results according to a damped sine function

\begin{equation}\label{7}
P(L)=P_{0}sin(k_{s}L-c)e^{-L/L_{s}}, 
\end{equation}

both the wave number $k_{s}$ and the spin-diffusion length $L_{s}$ can be estimated (see Table\ref{table1}). Thus, at energy $E_{1}$, the spin precesses with a period of $T_{s}=\hbar/k_{S} \alpha_{R} \approx 60 \ ps$, being $\alpha_{R}$ the Rashba strength for the AGNR which is extracted from reference.\cite{rojas2018} In addition, the spin-diffusion length reaches $\approx 17.2 \ \mu m$ which is one order of magnitude greater than previously reported values for AGNRs.\cite{salimath2014}

Similar analysis at energy $E_{2}$ (where the resonance is located) yields a value for the spin-diffusion length of $\approx 33.1 \ \mu m$, \emph{i.e.}, twice as large. This enhancement is consistent with\emph{spin hot spot} scenarios.\cite{kochan2014} Note that the amplitude of oscillation for this case is smaller than the previous one, also by a factor 2. The inset shows the oscillations for extrapolated longer length scales, both cases reflecting the oscillations.

\begin{table}[h]
\small
\caption{\small Estimated values of oscillation wave number, Rashba strength, spin-diffusion length, and period of oscillation of two different energies corresponding to valence and conduction bands.}
\begin{tabular}{ c c c c c }
 \hline
 Energy  & \parbox[t]{1.5cm}{$k_{s}$ ($nm^{-1}$)} & \parbox[t]{1.5cm}{$\alpha_{R}$ ($eV \AA$)} & \parbox[t]{1.5cm}{$L_{s}$ ($\mu m$)}  &  \parbox[t]{1.5cm}{$T_{s}$ ($ps$)} \\ 
\hline \\ 
\small   $E_{1}$     &     0.0023             &    0.300    &    17.2     &    59.9    \\
\small  $E_{2}$      &      0.012           &     0.300    &    33.1     &    11.5    \\
\hline
\end{tabular}
\label{table1}
\end{table}

\subsection{Concentration effects}
\begin{figure}[t]
\centering
\includegraphics[scale=0.38]{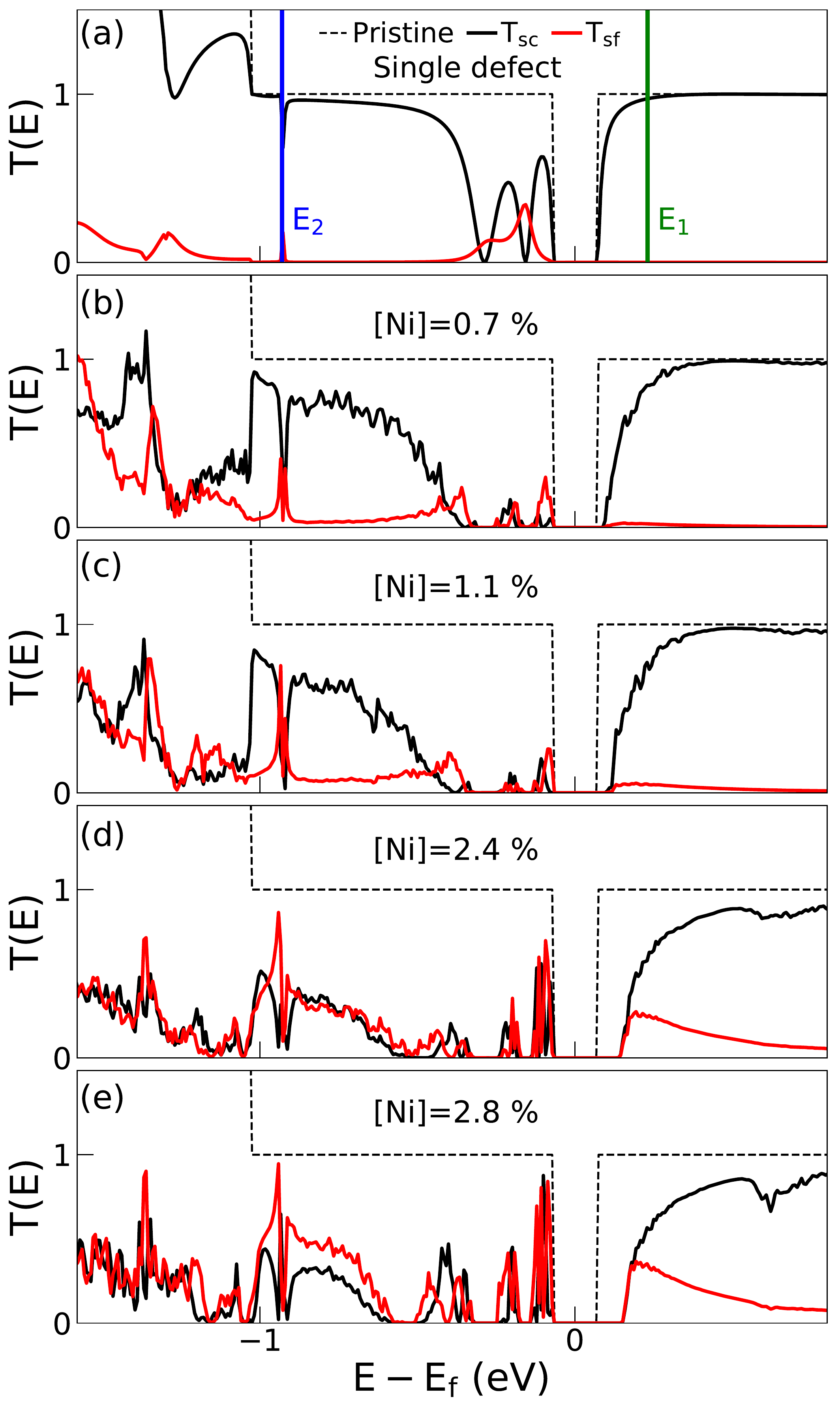}
\caption{Transmission probability for spin-conserved (black lines) and spin-flip (red lines) currents for four different Ni adatom concentrations. The ribbon lengths ($L$) is set up at $\approx 100 \ nm$. The single defect case is included for the sake of clarity of the resonance observed in the valence band. Dashed lines indicate the transmission of pristine AGNRs. Vertical lines spotlight energies $E_{1}$ and $E_{2}$ taken for the relaxation length analysis.}\label{fig4}
\end{figure}

\begin{figure}[t]
\centering
\includegraphics[scale=0.35]{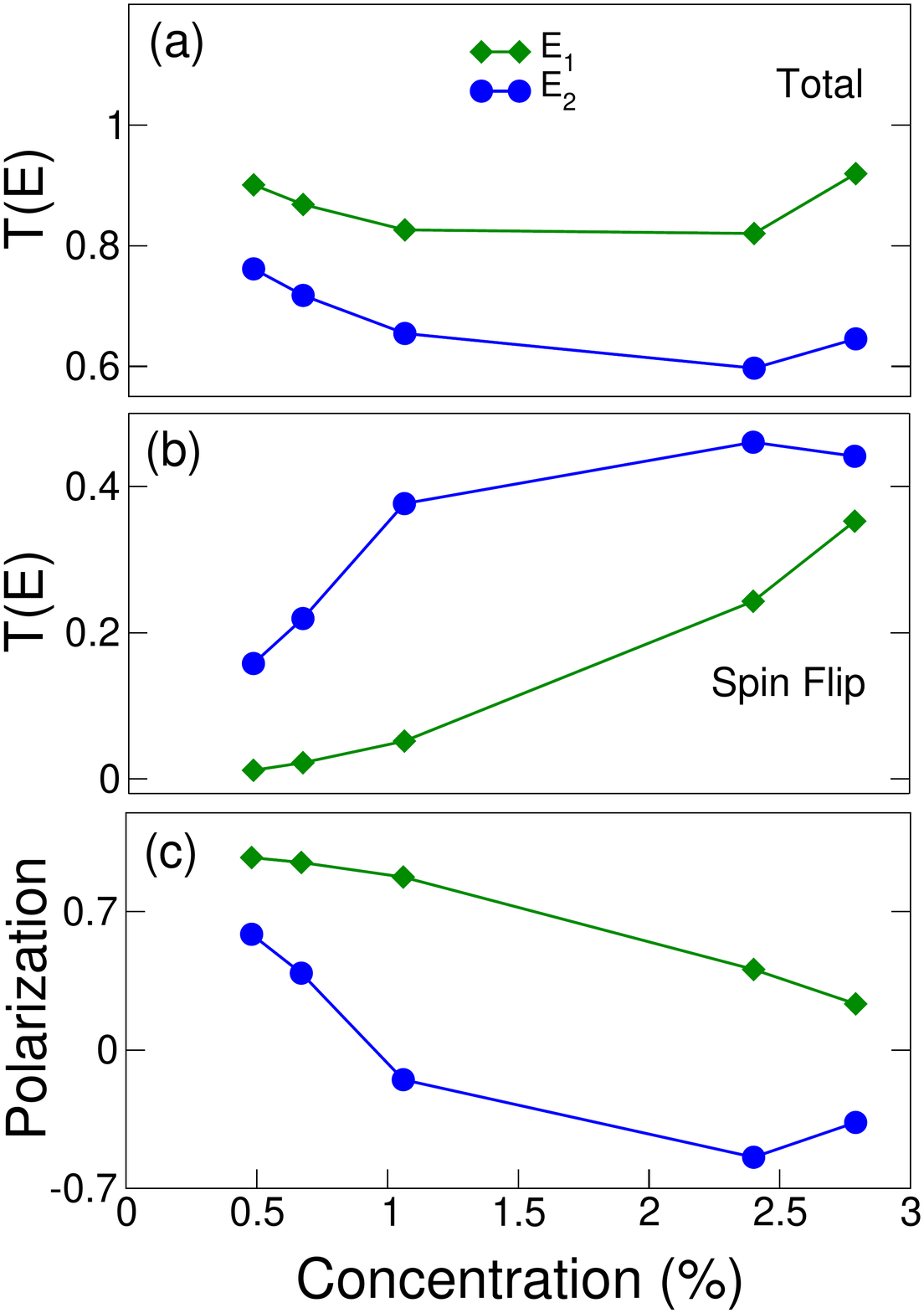}
\caption{Transmission probabilities as a function of concentration for energies $E_{1}$ and $E_{2}$ corresponding to conduction band (green squares) and valence band around the resonance (blue circles): (a) Total transmission probability, and (b) Spin-flip transmission probability. (c) Polarization as a function of concentration.} \label{fig5}
\end{figure}

In Figure \ref{fig4}(b-e) we present the transmission for different impurity concentration. We noted that for electronic states in the conduction band, the spin-conserved channel slowly decreases with concentration, and the appearance of a dip at higher concentrations becomes evident. For electronic states at valence band two trends are clearly depicted: i) around the Fermi level, the increase of scatter centers greatly enhances the spin-flip transmission probabilities and ii) for energies around the resonance, the spin-flip channel increases with the concentration becoming comparable with the spin-conserved component. Overall, as shown in Figure \ref{fig5}(a), we note that the total transmission for the selected energies follow the same trend.

In Figure \ref{fig5}b, and for $E_{1}$, we observe a monotonic behavior of the spin-flip channel with respect to the concentration. In addition, for the averaged energies around the resonance ($E_{2}$), we observe that for concentrations below $1 \%$ the channel transmission increases with concentration, while for values above this threshold, the transmission probability slowly decreases. This opposite trend might be seen as a signature of the predominance of other spin scattering mechanism regime which arise for moderate concentrations above $1 \%$.

Based on the polarization behavior limits $\lim_{L\to 0} P(L)$, where the amplitude of oscillation reach its maximum value and $\lim_{L\to\infty} P(L)$, in which the polarization must go to zero, it is possible to discuss the implications of our results to provide insights about the dominant spin relaxation mechanism in the studied system.

Therefore, by considering the oscillatory behavior of polarization as a function of length for a given concentration, as seen in Figure \ref{fig3}(a-b), one might expect that as the polarization decrease, the decay factor (inverse of spin relaxation length) should also be reduced. Consequently, as observed from Figure \ref{fig5}c, and considering electronic states in the conduction band ($E_{1}$), one might expect that the spin relaxation length increases at approximately linear rates with respect to concentration of impurities. This behavior is a signature of the Dyakonov-Perel (DP) mechanism similar to the observed in graphene on hBN substrate.\cite{tuan2016}

For energies close to the resonance ($E_{2}$), similar trends are observed specially for concentrations smaller than $1 \%$. Nevertheless, for concentrations above $1 \%$ an opposite trend of the polarization is observed, suggesting a different Physical origin that can be possibly related to the Elliot Yafet (EY) mechanism.\cite{Elliot1954,Yafet1952}

It is worth mentioning that our results slightly reveal the character of the dominant spin relaxation mechanism in GNRs indirectly. Thus, further theoretical and experimental works addressing directly  the relation between spin-lifetime and concentration are needed to unambiguously determine the nature of the mechanisms that rule the spin lifetime in graphene nanoribbon devices.

\section{Conclusions}
The spin-dependent transport properties of armchair graphene nanoribbons in the presence of extrinsic 
spin-orbit coupling induced by a random distribution of Nickel adatoms is studied. By combining the recursive Green's function formalism with density functional theory, we explore the influence of ribbon length and metal adatom concentration on the transmission probabilities. At a given length, we observed a significant enhancement of the spin-flip channel around resonances and at energies right above the Fermi level. Moreover, we estimate the spin-relaxation length finding values on the order of tens of micrometers at low adatom concentrations. This study is conducted at unprecedented ribbon lengths entirely carried out with fully \emph{ab-initio} methods which provide evidence of such relaxation length and the observation of oscillations in the spin-polarization. Finally, our results suggest that at lower concentration of Ni adatoms, the dominant spin relaxation mechanism resembles to the one proposed by Dyakonov-Perel (DP).

\section*{Conflicts of interest}
There are no conflicts to declare.

\begin{acknowledgments}
This research received the financial support from the Brazilian agencies CAPES, CNPq and FAPESP grant number 2012/24227-1. A.R.R. acknowledges support from ICTP-SAIRF (FAPESP project 2011/11973-4) and the ICTP-Simons Foundation Associate Scheme.  This work uses the computational resources from GRID-UNESP and CENAPAD/SP.
\end{acknowledgments}


%

\bibliography{apssamp}

\end{document}